\documentclass[useAMS,usenatbib]{mn2e}
\usepackage{epsfig}
\usepackage{amsmath}
\usepackage[authoryear]{natbib}
\usepackage{longtable}
\usepackage{graphicx}
\usepackage{amssymb}
\usepackage{url}
\usepackage{color}
\usepackage{grffile}
\usepackage{comment}
\usepackage{xspace}

\def\mh{\mbox{$M_{\rm halo}$}} 
\def\mstell{\mbox{$M_{\rm stell}$}} 
\def\msprim{\mbox{$M_{\rm stell,prim}$}} 
\def\mssec{\mbox{$M_{\rm stell,sec}$}} 

\def\mpc{$h^{-1}~\rm{ Mpc}$}

\def\msun{h^{-1}~\rm{ M_{\bigodot}}}

\newcommand{\gl}{{\sc galform}\xspace}
\newcommand{\lgal}{{\sc l-galaxies}\xspace}
\title[Galactic Conformity in SAMs]
  {Galactic conformity measured in semi-analytic models}
\author[Lacerna et al.]
  {I. Lacerna,$^{1, 2, 3}$\thanks{E-mail: ialacern@astro.puc.cl} S. Contreras,$^{2, 4}$ R. E. Gonz\'alez,$^{2, 4}$ N. Padilla$^{2, 4}$ and V. Gonzalez-Perez$^{5}$\\
    $^1$Instituto Milenio de Astrof\'isica, Av. Vicu\~na Mackenna 4860, Macul, Santiago, Chile\\
    $^2$Instituto de Astrof\'isica, Pontificia Universidad Cat\'olica de Chile, Av. V. Mackenna 4860, Santiago, Chile\\
    $^3$Astrophysical Research Consortium, Physics/Astronomy Building, Rm C319, 3910 15th Avenue NE, Seattle, WA 98195, USA\\
    $^4$Centro de Astro-Ingenier\'ia, Pontificia Universidad Cat\'olica de Chile,
Av. V. Mackenna 4860, Santiago, Chile\\
	$^5$Institute of Cosmology and Gravitation, University of Portsmouth, Dennis Sciama Building, Portsmouth PO1 3FX, UK\\    
}    

\date{}

\pagerange{\pageref{firstpage}--\pageref{lastpage}} \pubyear{000}

\begin{document}

\label{firstpage}

\maketitle
\begin{abstract} 
We study the correlation between the specific star formation rate of central
galaxies and neighbour galaxies, also known as `galactic conformity',  out to 20 \mpc\ using three semi-analytic models (SAMs, one from \lgal and other two from \gl).
The aim is to establish whether SAMs are able to show galactic conformity using different models and selection criteria.
In all the models, when the selection of primary galaxies is based on an isolation criterion in real space,
the mean fraction of quenched galaxies around quenched primary galaxies is higher than that around star-forming primary galaxies of the same stellar mass.  
The overall signal of conformity decreases when we remove satellites selected as primary galaxies, but the effect is much stronger in \gl models compared with the \lgal model.
We find this difference is partially explained by the fact that in \gl once a galaxy becomes
a satellite remains as such, 
whereas satellites can become centrals at a later time in \lgal.
The signal of conformity decreases  down to 60\% 
in the \lgal model after removing 
central galaxies that were ejected from their host halo in the past.
Galactic conformity is also influenced by primary galaxies at
fixed stellar mass that reside in dark matter haloes
of different masses.
Finally, we explore 
a proxy of
conformity between distinct haloes. 
In this case the conformity is weak beyond $\sim 3$ \mpc\ ($<$3\% in \lgal, $<$1--2\%  in
\gl models). Therefore, it seems
difficult that conformity is directly related
with a long-range effect. 
\end{abstract}
\section{Introduction}

The description of physical properties of galaxies with their environment is
paramount for understanding galaxy formation. 
A remarkable case is galactic conformity that is a term used to describe the observed correlation between star formation in central galaxies and in their neighbour galaxies.
\citet{Weinmann+2006} defined the term of galactic conformity after finding that quenched central galaxies have a higher fraction of quenched satellite galaxies compared to star-forming central galaxies in galaxy groups of similar mass at $z < 0.05$ 
in the New York University Value Added Catalogue \citep[NYU-VAGC,][]{Blanton+2005}, based on SDSS DR2 \citep{Abazajian:2004}. 
Later, \citet{Kauffmann+2013} found a galactic conformity effect
between low-mass central galaxies with low specific star formation rate (sSFR) or gas content and neighbour galaxies with low sSFR out to scales of 4 Mpc at $z < 0.03$
in SDSS DR7 \citep{Abazajian:2009}. 
These results motivated the distinction of the conformity measured at small separations between the central galaxy and their satellite galaxies within a dark matter halo as one-halo conformity, whereas the signal measured at large separations of several Mpc between the central galaxy and neighbour galaxies in adjacent haloes as two-halo conformity \citep{Campbell+2015,Hearin+2015}.

Galactic conformity has been measured both in the local Universe and at higher redshifts. 
In addition to the results of \citet{Weinmann+2006} and \citet{Kauffmann+2013}, 
\citet{WangWhite2012} found that red central galaxies have redder satellites than blue centrals of the same stellar mass using SDSS DR7 and SDSS DR8 \citep{Aihara+2011}.
\citet{Phillips+2014} found that satellites are more quenched around massive quenched galaxies compared to a control sample at fixed stellar mass from SDSS DR7. 
\citet{Knobel+2015} confirmed that satellites around quenched 
centrals are more likely to be environmentally quenched
than those around non-quenched centrals using galaxy groups from SDSS DR7 in the redshift range 0.01 $< z < 0.06$.
\citet{Tinker+2017} and \citet{Sin+2017} studied the two-halo conformity in SDSS DR7, and \citet{Kerscher2017} studied the range of conformity in SDSS DR12 \citep{Alam+2015}.
At higher redshifts,
\citet{Hartley+2015} found a tendency for passive satellites to be preferentially located around passive central galaxies using the UKIRT Infrared Deep Sky Survey 
\citep[UKIDSS;][]{Lawrence+2007}
Ultra Deep Survey (UDS) with photometric redshifts at $0.4$ $<$ $z$ $<1.6$.
\citet{Kawinwanichakij+2016} found that satellites around quiescent central galaxies
are more likely to be quenched compared to the satellites around star-forming centrals
from the deep near-infrared surveys ZFOURGE/CANDELS, UDS, and UltraVISTA \citep{McCracken+2012} with photometric redshifts  $0.3$ $<$ $z$ $<2.5$. They found that the significance of this one-halo conformity signal varies with redshift
($\gtrsim 3\sigma$ for $0.6 < z < 1.6$, whereas it is only weakly significant at $0.3 < z < 0.6$ and $1.6 < z < 2.5$).
\citet{Berti+2017} reported an excess of star-forming neighbours around star-forming central galaxies, of 
$\sim5\%$ on scales of $0-1$ Mpc and a two-halo signal of $\sim1\%$ on scales of $1-3$ Mpc using the PRism MUlti-object Survey 
\citep[PRIMUS;][]{Coil+2011,Cool+2013} with spectroscopic redshifts at $0.2$ $<$ $z$ $<$ 1.0. These signals are weaker than those detected at $z \lesssim 0.05$.

It is still a matter of debate why quenched central galaxies tend to reside preferentially in quenched environments out to several Mpc scales. 
\citet{Kauffmann2015} suggests that it could be related to AGN feedback that extends beyond the virial radius of massive haloes since they found an excess of massive neighbour galaxies hosting radio-loud AGN around low-mass, quenched central galaxies. 

\citet{Hearin+2016} studied the correlations between the mass accretion rates of nearby haloes as a potential physical origin for this effect. 
They found that pairs of host haloes may show correlated assembly histories even when their present-day separation is greater than thirty times the virial radius of either halo. Therefore, these authors suggest that galactic conformity is related to large-scale tidal fields and that its signal should decrease with redshift. If the star formation history of central galaxies is coupled with some host halo property, that could mean that the two-halo conformity is related to  the ``assembly bias" \citep{Hearin+2015}, where the large-scale clustering of dark matter haloes shows a dependence on halo properties beyond the halo mass \citep[e.g.][]{Gao05,
Wechsler06,Zhu06,Bett+2007,Gao-White07,Wetzel+2007,Li+2008,Angulo+2008b,FW10,Lacerna2011,Lacerna2012,vanDaalen+2012,Hearin2015,Lin+2016,More+2016,Tojeiro+2017,Zu+2017}.
In this context, \citet{Berti+2017} mention that their observational results are consistent with the  predictions of \citet{Hearin+2016} and that two-halo galactic conformity is reflecting assembly bias. 

However, \citet{Paranjape+2015} argued if the conformity measured by \citet{Kauffmann+2013} \citep[and probably that by][]{Berti+2017} at projected scales $\lesssim$ 4 Mpc in similar stellar mass bins is evidence of assembly bias. 
Based on results using 
a mock galaxy catalogue generated from a halo occupation distribution (HOD) function,
they say that only at very large separations ($\gtrsim$ 8 Mpc) there is a genuine two-halo conformity that is driven by the assembly bias of small host haloes.  
They suggest that the observed conformity at $\lesssim$ 4 Mpc
is just because of central galaxies with the same stellar mass can be residing in host haloes of different halo masses.
\citet{Tinker+2017} reproduced the result of \citet{Kauffmann+2013}, but they have shown it is mainly driven by contamination in the isolation criterion to select the sample of central galaxies 
(see also \citealt[][]{ZM+2017}). 
After removing a small fraction of satellite galaxies, they detect only a small conformity signal out to projected distances of 2 Mpc. They suggest that $\sim 2-5\%$ differences in the quenched fractions of neighbour galaxies at projected distances between 1 and 3 Mpc can be produced by mechanisms other than halo assembly bias.
\citet{Sin+2017} studied in detail the conformity signal measured by \citet{Kauffmann+2013}. In addition to the misclassification of satellite galaxies as centrals in the isolation criterion, they mention that this signal is strongly amplified by weighting in favour of central galaxies in very high-density regions, and the use of medians to characterize the bimodal distribution of sSFR. They conclude that the large-scale conformity presented in  \citet{Kauffmann+2013} is a relatively short-range effect that originates
from a very small number of central galaxies in the vicinity of just a few very massive clusters, rather than a very long-range effect.

Interestingly, \citet{Kauffmann+2013} and also \citet{Kauffmann2015} found a much weaker galactic conformity in synthetic models (semi-analytic models and hydrodynamical simulations) compared to observations. 
This indicates that there could be physical processes operating in galaxies that are not included in these models.
However, \citet{WangWhite2012} mention that the semi-analytic model of \citet{Guo:2011} reproduces qualitatively the trends observed between central galaxies and satellites within a projected distance of 300 kpc.
Similarly, \citet{Sin+2017} have recently found that the semi-analytic models of \citet{Henriques:2015} and \citet{Guo:2011} show
similar results in terms of both the amplitude and the range of the sSFR correlation to their observational results.
Furthermore, \citet{Bray+2016} claim to have measured a strong signal of galactic conformity using the Illustris simulation \citep[][]{Vogelsberger+2014}, which is a large hydrodynamical simulation. They find that the mean red fraction of galaxies around redder neighbour galaxies is higher than around bluer galaxies at fixed stellar mass out to distances of 10 Mpc. These authors conclude that the measured amplitude of the conformity signal depends on the criteria to select central galaxies in observations, projection effects, and stacking techniques.
\citet{RD+2017} have found that galaxy conformity is evident in the MUFASA hydrodynamical simulation \citep{Dave+2016} using different tracers (sSFR, colour, and atomic gas), but declines rapidly at scales larger than 1 Mpc.

The aim of this paper is to measure the signal of galactic conformity using three different semi-analytic models (SAMs) as it has been suggested that this type of numerical models may not show the strong signal of galactic conformity measured in observations (e.g. \citealt{Kauffmann+2013}). Nevertheless, our goal is not to reproduce or compare with the observational signal, but to establish whether 
SAMs in general 
are able to show a strong signal
for this effect using different models and selection criteria.
In addition, we discuss our results 
in different scale ranges
so that we can obtain a distinction 
among the models
of the conformity measured within small separations from the central galaxy (one-halo conformity) and, especially, at large separations 
(two-halo conformity) where the signal of conformity is still in debate. 

The outline of the paper is as follows. The descriptions of the three semi-analytic models used in this paper are presented in Section \ref{sec_data}. We define the samples and the method to measure the galactic conformity in Section \ref{sec_definitions}. 
The results are shown in Section \ref{sec_results}.
Here we point out the main differences
among the three SAMs, and explore the theoretical origin of those differences.
The implications of our results 
with a particular emphasis in the
existence of two-halo conformity
are discussed in Section \ref{sec_discussion}. 
Finally, our conclusions are given in Section \ref{conclusions}.

Throughout this paper we use the reduced Hubble constant $h$,
where $H_0 = 100$ $h$ km s$^{-1}$ Mpc$^{-1}$, with the following dependencies: stellar mass and halo mass in $\msun$, physical scale in \mpc, and the specific star formation rate in $h$ yr$^{-1}$,
unless the explicit value of $h$ is specified.

\section{Data}
\label{sec_data}

The objective of this paper is to test if semi-analytic models of galaxy formation 
are able to show galaxy conformity. 
In this section we introduce the three galaxy formation models used in this work. The models used in this paper have different implementations of 
the physical processes involved in galaxy evolution.
By comparing models from different working groups we can get insight for which predictions are robust and which depend on the particular implementation of the physics.

\subsection{The semi-analytic models}
\label{sec_SAM}

Semi-analytical models (SAMs, e.g. \citealt{Cole:2000}) of galaxy formation establish a physically motivated model of how galaxies form and evolve within a cosmological context. Some of the physical processes that are modeled are
the shock heating and radiative cooling of gas inside dark 
matter haloes that lead to the formation of galactic discs, the feedback from supernovae (SNe), from the accretion of mass onto super-massive black holes and from photoionization heating of the intergalactic medium (IGM), chemical enrichment of the stars and gas, 
and galaxy mergers driven by dynamical friction within dark matter haloes, which lead to the formation of stellar spheroids and also may trigger bursts of star formation. 

In this work we use three SAMs:  \citet[][hereafter G13]{Guo:2013} that is a version \lgal, developed by the Munich group \citep{DeLucia:2004, Croton:2006, DeLucia:2007, Guo:2011, Henriques:2013,Henriques:2015}, and two flavours of \gl \footnote{The output of GP14 and G13 models are publicly available from the Millennium Archive in Durham 
\url{http://virgo.dur.ac.uk/data.php} and in Garching \url{http://gavo.mpa-garching.mpg.de/Millennium/}.}, developed by the Durham group \citep{Bower:2006, Font:2008, Lagos:2012, Lacey:2016}. The two \gl flavours are \citet[][hereafter GP14]{Gonzalez-Perez:2014} and that same model but with a gradual ram-pressure (GRP) stripping of the hot gas in satellite galaxies, instead of the default instantaneous stripping (see \citealt[][]{Lagos:2014b} for a comparison between GP14 and GRP).

The three models used in this study were run over the same dark matter simulation, the
MS-W7 simulation \citep{Guo:2013,Jiang:2014,Gonzalez-Perez:2014}. Similar to the Millennium simulation \citep{Springel+2005}, the MS-W7 simulation contains $2160^3$ particles in a periodic box of 500 $h^{-1}$ Mpc on a side, but with a mass resolution of 9.3 $\times$ $10^8$ $\msun$ and with a WMAP7 cosmology\footnote{The values of the cosmological parameter of this simulation are: $\Omega_{m0}$  = 0.272, $\Omega_{\Lambda0}$  = 0.728, $\Omega_{b0}$  = 0.0455, $\sigma_8$ = 0.81, $n_{\rm s}$ = 0.967 and $h$ = 0.704} \citep[][]{Komatsu+2011}.

An overview of the three models is provided below, focused on the aspects where they differ and that are relevant for this study \citep[see also][]{Contreras+2013,Guo+2016}.

\subsubsection{Merger trees and haloes hosting central galaxies}

All the models use a Friends-of-Friends group finding algorithm \citep[FOF,][]{Davis+1985} to identify haloes in each snapshot of the simulation that contain at least 20 particles, and then they run SUBFIND on these groups to identify subhaloes within the FOF groups \citep{Springel+2001}. The models use different merger tree algorithms: G13 builds the merger trees following \citet{Springel2005}, \citet{DeLucia:2007} and \citet{Boylan-Kolchin+2009} and GP14 and GRP use the {\small DHALO} algorithm \citep{Jiang:2014}. Both methods identify descendants of a halo at the following 2 snapshots in the case of \lgal and 5 in the Dhalo algorithm. In both cases there is an attempt to avoid the premature link of haloes which pass through another halo, but the {\small DHALO} algorithm also imposes that the halo mass increases monotonically 
with the age of the universe \citep{Jiang:2014}. Thus, the haloes identified as such and their mases are different between the \lgal and \gl models, however the differences are small enough to not affect the conclusions of this work 
(for further details on this point see \citealt{Jiang:2014} and \citealt{Guo+2016}).

Central galaxies in \lgal are hosted by the most massive subhalo (main subhalo) within a FoF group. \gl defines host haloes either at the last time step of the simulation or at the last time a galaxy was a classified as central and determines its main progenitor as the one that contributed the most bound particles. In \gl once a galaxy becomes a satellite, it will be treated as such until the end of the simulation. This is not the case in \lgal, in which satellites beyond the virial radius of the halo that was hosting them can become centrals again. Note that this is not just a matter of reclassification, as the gas accretion and stripping is modeled differently for central and satellite galaxies.

\subsubsection{Gas in satellite galaxies}

As it has been previously mentioned, the only difference between the GP14 and GRP models is the way gas is stripped from satellite galaxies due to ram-pressure. In GP14, a galaxy is assumed to lose its hot gas completely and instantaneously once it becomes a satellite; in both GRP and G13, this process is gradual and depends on the orbit of the satellite. The G13 model also allows for tidal stripping of gas. The stripping of gas in satellite galaxies affects the conformity prediction at small scales (i.e. one-halo conformity), where in GRP and G13 galaxies will become redder earlier, but should not affect the results at large scales (i.e. two-halo conformity).

\subsubsection{Star formation and feedback}

For the quiescent formation of stars in discs, the G13 model assumes that, once the surface gas density exceeds a critical value, stars will form following a simplified empirical Kennicutt relation \citep{ken98}. In \gl, the quiescent star formation in galaxy discs
explicitly depends on the molecular component of the gas, with the star formation rate being proportional to the surface density of molecular hydrogen in the ISM \citep{Lagos:2011a}. Note that the empirically motivated calculation followed in \gl does not impose a threshold for the star formation to happen.

The G13 model assumes the bursts of star formation to be proportional to the total mass of cold gas and the mass ratio of two merger progenitors whenever a merger happens. \gl assumes that bursts of star formation are simply proportional to the total mass of cold gas present in galaxy bulges and inversely proportional to a star formation time-scale \citep{granato00}. 

In the three models, when massive stars die, cold gas is ejected from galaxies at a rate proportional to the SFR, with the proportionally factor depending on the circular velocity of the galaxy. This dependency is assumed to be a power law in \gl, while the G13 model assumes a more complicated form. In both models, the gas that is ejected from the halo will be gradually reincorporated at times that are different between \lgal and \gl. See \citet{Guo+2016} for further details on how these physical processes compare between \lgal and \gl.

\section{Definitions}
\label{sec_definitions}

\subsection{Primary galaxies}
\label{sec_prim}

In this paper, we estimate the fraction of quenched secondary galaxies around primary galaxies. The latter correspond to the most massive or the brightest galaxy in a group (or halo). By following a similar observational approach of \citet[][see also \citealp{Kauffmann2015}]{Kauffmann+2013}, 
we define a galaxy with stellar mass \mstell\ to be a primary galaxy if there
is no other galaxy with stellar mass greater than \mstell/2 within a given radius. In practice, this is an isolation criterion where we use the 3D positions of galaxies taking into account the periodic boundary conditions in the simulation box.
We use three isolation radii in real space: 0.1, 0.5 and 1 \mpc.

\subsection{Secondary galaxies}
\label{secondaries}

The secondary galaxies correspond to all the galaxies in the vicinity of primary galaxies. The minimum stellar mass for the secondaries is $10^9$ $\msun$. Recall that due to the isolation criterion, the maximum stellar mass for the secondary galaxies is \mstell/2 of the primary galaxy (\mssec\ $<$ \msprim/2) inside the isolation radius.

\subsection{Galactic conformity}
\label{sec_GC}

\begin{figure*}
\epsfysize=5.8cm \epsfbox{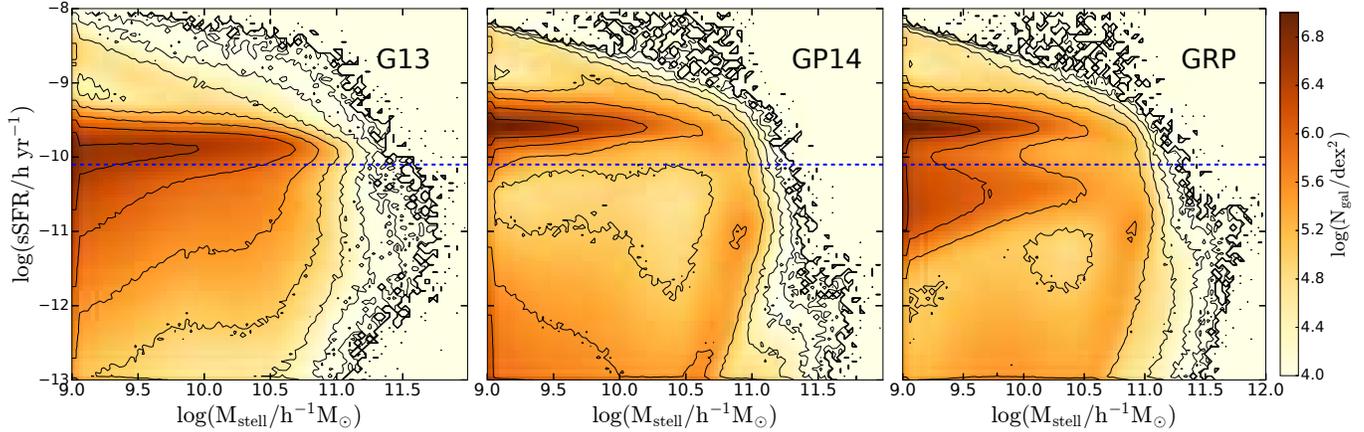}
\caption{
Contours with the distribution of sSFR as a function of stellar mass for the synthetic galaxies of G13 (left-hand panel), GP14 (middle panel), and GRP (right-hand panel).
The blue dashed horizontal line is the fiducial value used in this work that roughly separates star-forming and quenched galaxies (above and below the line, respectively). The colour bar indicates the number density of galaxies.
}
\label{plot_sSFR}
\end{figure*}

In order to detect galactic conformity in the semi-analytic models, we measure the mean fraction of quenched secondary galaxies 
around quenched (Q) and star-forming (SF) primary galaxies. 
We define that there is a galactic conformity signal when the mean quenched fraction of secondaries at a given distance from Q primary galaxies is higher than that around SF primary galaxies of the same stellar mass. 
Fig. \ref{plot_sSFR} shows the distribution of the specific star formation rate (sSFR, simply defined as the star formation rate normalized by the stellar mass) for the synthetic galaxies of G13, GP14 and GRP as a function of stellar mass. 
We refer to a galaxy as ``quenched" if the sSFR is lower or equal than 10$^{-10.1}$ $h$ yr$^{-1}$. Otherwise, the galaxy is considered as star forming. This value of sSFR is marked as a blue dashed line in Fig. \ref{plot_sSFR}. 
We obtain that the fraction of Q galaxies is similar for the three models using this cut in sSFR (57\% for G13, 58\% for GP14, and 58\% for GRP). We have checked that our results are robust changing this cut 
by $0.3$ and $0.5$ dex.

The errors in the estimation of the mean quenched fractions are calculated
using the jackknife method \citep[e.g.][]{Zehavi02,Norberg09}. In detail, we split every sample in 120 subsamples and estimate the covariance matrix as follows
\begin{eqnarray}
C(p_i,p_j) = \frac{N-1}{N}\sum_{k=1}^N(p_{i}^k - \overline{p}_i)(p_j^k - \overline{p}_j),
\label{eq_C}
\end{eqnarray}
where $N$ is the number of subsamples, $p_i$ is the mean quenched fraction around primary galaxies in the $i$th radial bin, and 
\begin{eqnarray}
\overline{p}_i = \frac{1}{N}\sum_{k=1}^N p_{i}^k.
\end{eqnarray}
Error bars in the figures of Section \ref{sec_results} are estimated using the diagonal of the covariance matrix, i.e., $\sqrt{C(p_i,p_i)}$.

We use the covariance matrix to estimate the cumulative signal-to-noise ratio (S/N) of galactic conformity. 
The cumulative S/N at a distance $r$ is measured through the difference between the quenched fraction of secondary galaxies around Q and SF primary galaxies as follows
\begin{eqnarray}
(\textrm{S/N})^2 =  B^T A^{-1} B ,
\label{eq_SNR}
\end{eqnarray}
where $B = p_{Q}(\le r) - p_{SF}(\le r)$. Here, $p_{Q}$ and $p_{SF}$ are the mean quenched fraction of secondaries around Q primary galaxies and SF primary galaxies, respectively, considering the radial bins out to $r$. Furthermore, 
\begin{eqnarray}
A^{-1} = (C_{Q} + C_{SF})^{-1} ,
\end{eqnarray}
where $C_{Q}$ ($C_{SF}$) is the covariance matrix defined in equation (\ref{eq_C}) for the mean quenched fraction around the Q (SF) primary galaxies. In order to differentiate the signal of conformity at small and large scales, we estimate equation (\ref{eq_SNR}) separately for the one-halo conformity and two-halo conformity 
out to $r\sim 0.5$ \mpc\ and in the range $0.5 < r < 20$ \mpc, respectively.

\section{Results}
\label{sec_results}

\subsection{Isolated primary galaxies}
\label{sec_results_isoprim}

\begin{figure*}
\epsfysize=17.5cm \epsfbox{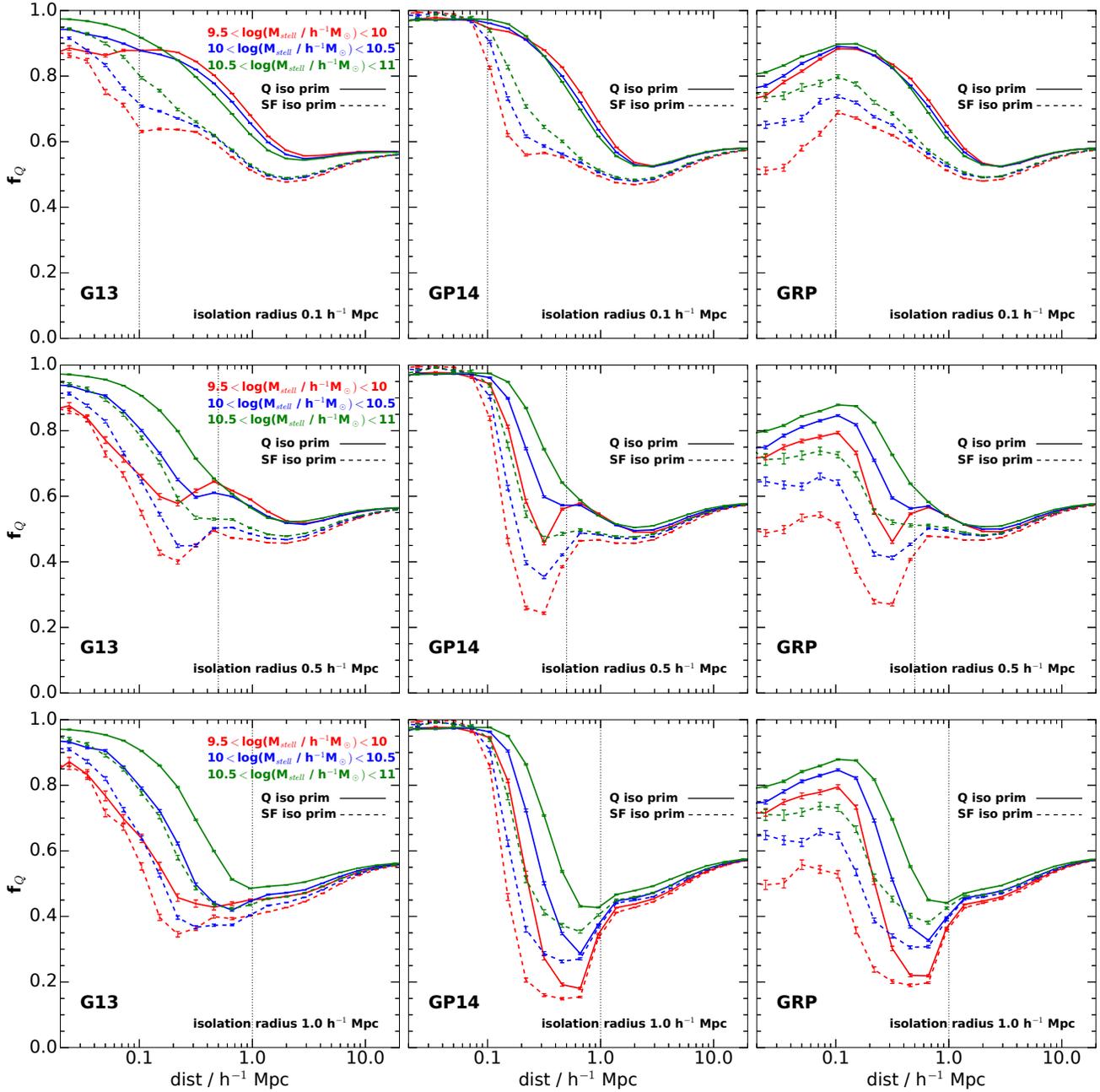}
\caption{
Mean quenched fraction of secondary galaxies as a function of the distance around quenched (Q, solid lines) and star-forming (SF, dashed lines) primary galaxies using the SAMs of G13 (left panels), GP14 (central panels), and GRP (right panels). The colours correspond to different stellar mass bins of primary galaxies indicated in the legend.  Primary galaxies are selected according to an isolation criterion in real space (see details in Section \ref{sec_prim}) with a radius of 0.1 \mpc\ (top panels), 0.5 \mpc\ (middle panels) and 1 \mpc\ (bottom panels) shown as a dotted-vertical line in each panel. The galactic conformity can be seen as the separation between the solid and dashed lines for a given mass sample (i.e. for the same colour lines).
}
\label{plot_Prim_iso_Sec_none}
\end{figure*}

Figure \ref{plot_Prim_iso_Sec_none} shows the mean quenched fraction of secondary galaxies as a function of the distance from primary galaxies in different stellar mass bins. Here the primary galaxies are selected according to the isolation criterion described in Section \ref{sec_prim} with a radius of 0.1 \mpc\ (top panels), 0.5 \mpc\ (middle panels) and 1 \mpc\ (bottom panels). 
The mean quenched fraction around Q primaries are shown in solid lines and that around SF primaries are shown in dashed lines. The galactic conformity can be seen as the separation between the solid and dashed lines for a given mass sample (i.e. for the same colour lines).

At small scales ($\lesssim 0.2$ \mpc), we detect galactic conformity in the G13 model (left-hand panels), i.e. the mean quenched fraction of secondary galaxies is higher around Q primary galaxies than that around star-forming (SF) primary galaxies of the same mass. The signal is present for the three stellar mass ranges of primary galaxies. On the other hand, the middle panels show 
the GP14 model with a full quenched population for scales below 0.1 \mpc\ caused by 
removing all hot gas in satellites
implemented in this model, making impossible to detect galactic conformity at these scales.
The GRP model (right-hand panels) shows the stronger galactic conformity at $\lesssim 0.2$ \mpc, i.e. larger separations between the solid and dashed lines, especially for the low-mass primaries (red lines). 

In general, the mean quenched fraction of secondaries decreases to larger distances out to some scale that is strongly dependent on the isolation radius of primary galaxies. 
At this radius, 
there is an upturn of the quenched fraction of secondary galaxies. For example, we can see this upturn at $\sim 0.1$ \mpc, $\sim 0.5$ \mpc, and $\sim 1$ \mpc\ in the top-right, middle-right and bottom-right panels, respectively. 
Recall that by definition, the secondaries have at most a half of the stellar mass of the primary galaxy inside the isolation radius.
Thus, it is more likely to find quenched secondary galaxies at and beyond this scale
since  the restriction on stellar mass is no longer present from the isolation criteria and, therefore, the mean quenched fraction of secondaries increases. 
The upturn in the quenched fraction is higher for low-mass primary galaxies (red lines) because the number of secondary galaxies with stellar masses \mssec\ $\geq$ \msprim/2 is higher compared with the case of more massive primaries.

We also detect the two-halo  galactic conformity ($\gtrsim$ 1 \mpc), but decreases to larger separations from the primary galaxies. This conformity (i.e. the separation between the solid and dashed lines) is lower compared to that in the one-halo conformity and depends on the stellar mass of the primary galaxy. The galactic conformity decreases in the G13 model  
as we increase the mass of the primary galaxies.
At distances of $\sim 10$ \mpc, there is no galactic conformity regardless the stellar mass and isolation radius, i.e. the mean quenched fractions of secondary galaxies around Q primary galaxies and SF primary galaxies at fixed stellar mass are almost the same. The galactic conformity at large-scales look to be smaller in the GP14 and GRP models compared to the G13 model.

Note that the mean quenched fraction of secondaries slightly increases for scales larger than 3 \mpc\ in all the SAMs because of quenched satellites in adjacent haloes.  

\begin{figure*}
\centering
\epsfysize=12cm \epsfbox{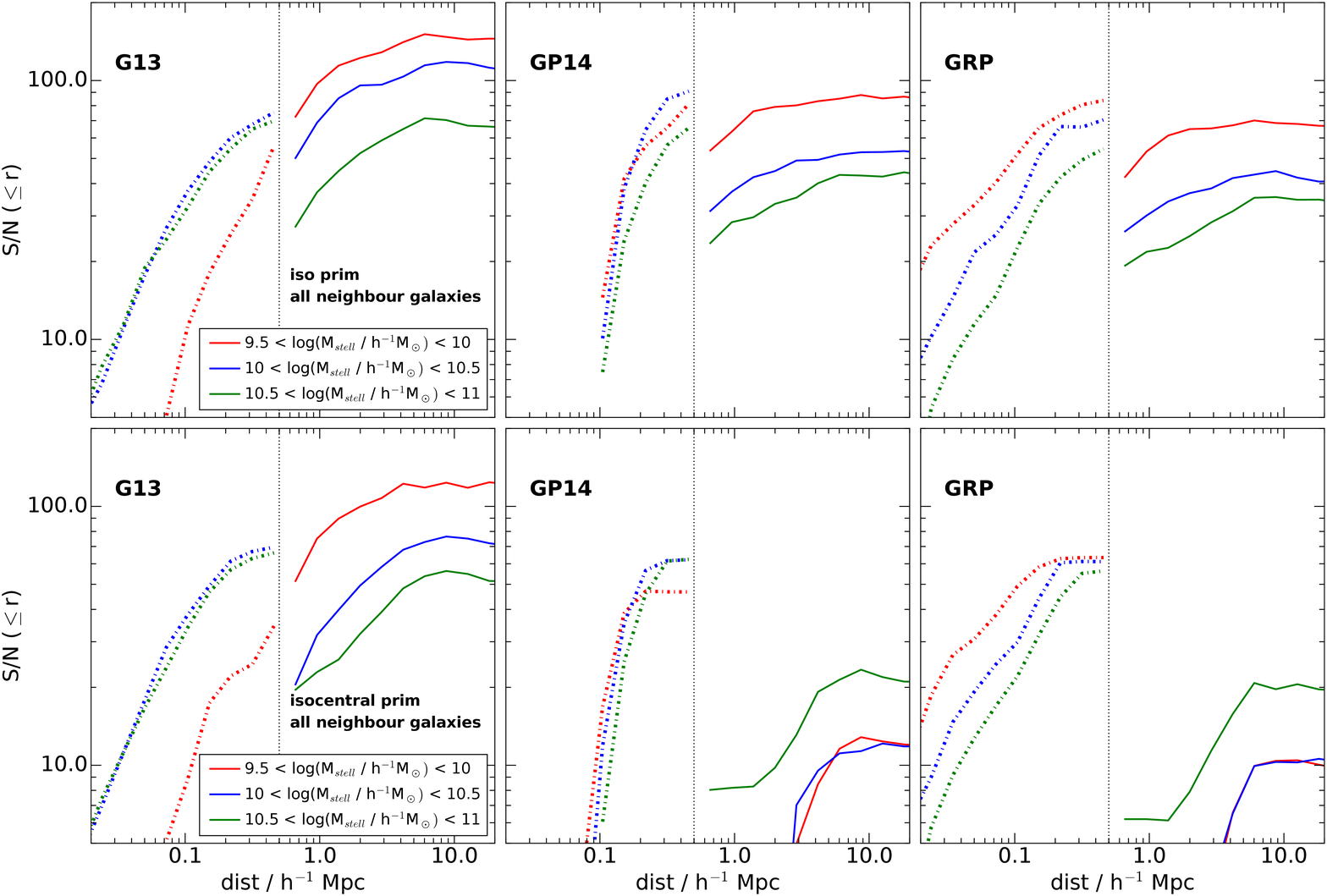}
\caption{Cumulative signal-to-noise ratio of galactic conformity 
using the SAMs of G13 (left panels), GP14 (central panels), and GRP (right panels).
The difference between the quenched fraction of secondary galaxies around Q and SF primary galaxies with the same stellar mass along with the respective 
covariance matrix are used to estimate the cumulative S/N (equation \ref{eq_SNR}).  
This is measured separately for the one-halo conformity out to $\sim 0.5$ \mpc\ (dash-dot lines) and the two-halo conformity in the range $0.5 < \textrm{dist} < 20$ \mpc\ (solid lines).
The isolation radius is shown as a dotted-vertical line in each panel, which also illustrates our scale to separate the one-halo conformity and the two-halo conformity.
In the top panels, the primary galaxies are selected according to an isolation criterion in real space with a radius of $0.5$ \mpc, whereas in the bottom panels 
they are also central galaxies in the simulations.
The colours correspond to different stellar mass bins of primary galaxies indicated in the legend.
}
\label{plot_SNR}
\end{figure*}

The top panels of Fig. \ref{plot_SNR} show the cumulative S/N (see Section \ref{sec_GC}) of galactic conformity for the cases with an isolation radius of 0.5 \mpc.
Hereafter, we only use this isolation radius  
since it is representative of selecting primaries (i.e. the overall results at Mpc scales do not change using a different isolation radius). 
The cumulative S/N is estimated as the difference in the quenched fraction of all the neighbour galaxies around Q and SF isolated primaries using equation (\ref{eq_SNR}). 
We estimate the cumulative S/N separately for the one-halo conformity out to $r\sim 0.5$ \mpc\ and the two-halo conformity in the range $0.5 < r < 20$ \mpc. 
Since this is a cumulative measurement with the scale, in this figure the cumulative S/N increases for scales where galaxy conformity signal is strong, while the cumulative S/N is relatively flat for scales where galaxy conformity signal is rather low.
In G13 (top-left panel of Fig. \ref{plot_SNR}), the one-halo
galactic conformity is stronger for primary galaxies more massive than $\msprim >$ $10^{10}$ $\msun$.
At larger scales, the galactic conformity becomes stronger between low-mass primary galaxies ($10^{9.5} -10^{10}$ $\msun$) and their neighbours.
Table \ref{table_iso} also shows that the cumulative signal-to-noise ratio of this case at $\sim$20 \mpc\ (third column) is higher compared to the other mass ranges of primary galaxies. 
The cumulative S/N of two-halo conformity is 2.6 times higher than the cumulative S/N of one-halo conformity for low-mass primaries, whereas the signal is very similar in both regimes for massive primary galaxies.
The last column of Table \ref{table_iso} 
corresponds to the ratio between the mean quenched fraction of secondaries around Q and SF isolated primaries ($f_{Q,Q prim}/f_{Q,SF prim}$) at 5 \mpc. This scale is used just as a reference of the two-halo conformity regime. 
$f_{Q}$ around low-mass Q primary galaxies is 7\% higher than that around SF primaries of the same mass at 5 \mpc, 
whereas it is 5\% between Q and SF primaries of higher masses.

The two-halo conformity of the GP14 and GRP simulations (top-middle and top-right panels of Fig. \ref{plot_SNR}) are qualitatively similar to that of G13. That is, the highest cumulative S/N is reached in the case between low-mass primary galaxies and their neighbour galaxies.
However, the results are quantitatively different. For a given mass range of primaries,  the cumulative S/N at 20 \mpc\ in GP14 and GRP are smaller compared with G13. 
Furthermore,
$f_{Q,Q prim}/f_{Q,SF prim}$ is also smaller for all masses (less than 4\%, see Table \ref{table_iso}).
In contrast to G13, the signal of two-halo conformity is in general not higher than that of one-halo conformity in both GP14 and GRP simulations.

\begin{table}
\caption{Conformity signal-to-noise ratio between isolated primary galaxies and secondaries (top panels of Fig. \ref{plot_SNR}). First column shows the mass range for the primary galaxies, 
the second column is the cumulative S/N out to 0.5 \mpc, and the third column is
the cumulative S/N measured from 0.5 \mpc\ out to 20 \mpc, S/N($0.5<r<20$). The last column
is the ratio between the mean quenched fraction of secondaries around Q and SF isolated primaries ($f_{Q,Q prim}/f_{Q,SF prim}$) at 5 \mpc.
The stellar mass is log$_{10}$ in units of $\msun$. 
}
\begin{tabular}{c c c c}
\hline
\hline
log(\mstell) & $\rm S/N$($r<0.5$) & $\rm S/N$($0.5<r< 20$) & ratio $f_Q(5)$   \\
\hline
G13 & & & \\
\hline
9.5-10  &  55.9  &  145.2  &  1.068 \\
10-10.5  &  75.3  &  112.0  &  1.051 \\
10.5-11   &  69.8  &  66.4  &  1.052 \\
\hline
GP14 & & & \\
\hline
9.5-10  &  81.2  &  86.6  &  1.030 \\
10-10.5  &  90.8  &  53.4  &  1.026 \\
10.5-11   &  65.6  &  44.2  &  1.037 \\
\hline
GRP & & & \\
\hline
9.5-10  &  83.8  &  66.9  &  1.024 \\
10-10.5  &  70.5  &  40.7  &  1.022 \\
10.5-11   &  54.5  &  34.7  &  1.032 \\
\hline
\end{tabular}
\label{table_iso}
\end{table}

\subsection{Isolated, central primary galaxies}

\begin{figure*}
\epsfysize=6.3cm 
\epsfbox{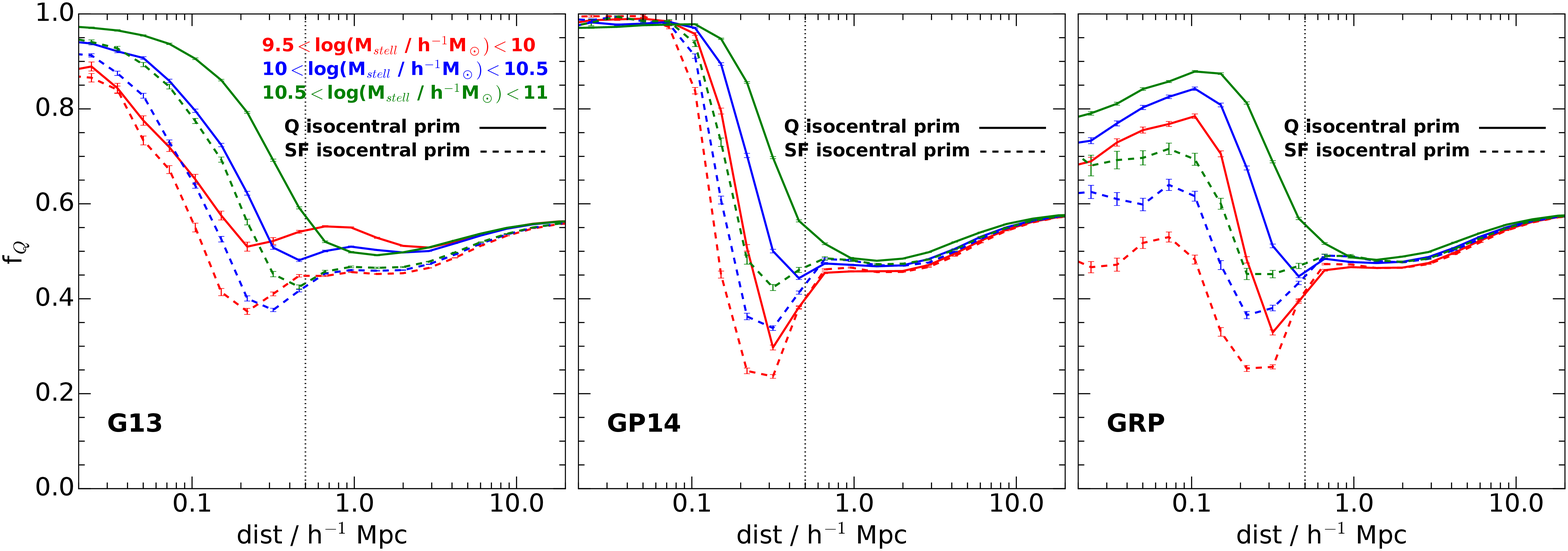}
\caption{
Mean quenched fraction of secondary galaxies as a function of the distance around Q (solid lines) and SF (dashed lines) primary galaxies using the SAMs of G13 (left-hand panels), GP14 (middle panels) and GRP (right-hand panels). Here the primary galaxies are selected according to an isolation criterion in real space with a radius of 0.5 \mpc\ 
shown as a dotted-vertical line
and, in addition, they are 
central galaxies in the simulations.
The colours correspond to different stellar mass bins of primary galaxies indicated in the legend.
}
\label{plot_Prim_IsoCentral}
\end{figure*}

Figure \ref{plot_Prim_IsoCentral} shows the case of isolated, central primaries with an isolation radius of 0.5 \mpc. Therefore, in addition of the isolation criterion used in observations for selecting central galaxies, they are actually central galaxies according to the numerical simulations. 
The results are similar to those using isolated primaries (middle row panels of Fig. \ref{plot_Prim_iso_Sec_none}) inside the isolation radius. However, it is clear that the galactic conformity is weaker at Mpc scales.

The bottom panels of Fig. \ref{plot_SNR} show that the cumulative S/N of two-halo conformity between the isolated, central galaxies and their neighbours is lower compared to that between isolated primaries and their neighbour galaxies (top panels) for the three SAMs. This means that part of the two-halo conformity detection
is polluted by satellites that are classified as primaries using only an isolation criterion for selecting central galaxies (see \citealt{Kauffmann+2013}, \citealt{Bray+2016}, and \citealt{Tinker+2017} for the same conclusion). 
We find that the fraction
of the low-mass primary galaxies which are actually central galaxies using an isolation radius of 0.5 \mpc\ is of
96\%  for G13, 95\% for GP14, and 94\% for GRP.

\begin{table}
\caption{
Same as Table \ref{table_iso}, but for isolated, central galaxies instead of using only isolated galaxies as primaries.
}
\begin{tabular}{c c c c}
\hline
\hline
log(\mstell) & $\rm S/N$($r<0.5$) & $\rm S/N$($0.5<r<20$) & ratio $f_Q(5)$\\
\hline
G13 & & & \\
\hline
9.5-10  &  34.5  &  123.9  &  1.060 \\
10-10.5  &  69.6  &  72.0  &  1.041 \\
10.5-11   &  66.1  &  51.5  &  1.044 \\
\hline
GP14 & & & \\
\hline
9.5-10  &  46.7  &  12.0  &  1.009 \\
10-10.5  &  62.2  &  11.8  &  1.010 \\
10.5-11   &  62.5  &  21.0  &  1.026 \\
\hline
GRP & & & \\
\hline
9.5-10  &  63.4  &  10.1  &  1.007 \\
10-10.5  &  61.2  &  10.6  &  1.009 \\
10.5-11   &  56.3  &  19.6  &  1.024 \\
\hline
\end{tabular}
\label{table_isocen}
\end{table}

For G13, we still find a consistent two-halo conformity signal between isolated, central primaries and secondaries (see also Table \ref{table_isocen}), although 
the cumulative S/N at 20 \mpc\ is lower when we include central galaxies in the primary sample compared with the case of only isolated primaries. In addition, $f_{Q,Q prim}/f_{Q,SF prim}$ at 5 \mpc\ is slightly smaller for all  masses (6\% for low-mass primaries, 4\% for primaries of higher masses).

However, for GP14 and GRP we do not find a strong two-halo conformity signal between isolated, central primaries and their neighbour galaxies. 
The cumulative S/N is $\sim$7 times smaller using central galaxies in the sample of low-mass primaries, $\sim$4 times smaller for primaries of intermediate masses, and $\sim$2 times smaller for massive primaries.
Furthermore, the ratio $f_{Q,Q prim}/f_{Q,SF prim}$ at 5 \mpc{} decreased to $\leq$1\% for primaries of \mstell $< 10^{10.5}$ $\msun$, and 2--3\% for massive primaries in GP14 and GRP.
Therefore, using central galaxies as primaries remarkably decreases the two-halo conformity compared to that using only isolated galaxies, but this strongly depends on %
each model. 
The differences in the treatment of gas in satellites among the models are not responsible for the differences measured in the two-halo galactic conformity between G13 and both GP14 and GRP.
We have confirmed that the results are qualitatively similar using only massive galaxies above \mstell\ $> 10^{10.5}$ $\msun$ where the sSFR distributions in Fig. \ref{plot_sSFR} are similar.

As mentioned in Section \ref{sec_SAM}, 
in G13 satellite galaxies that are far enough from the virial radius of the host halo in some snapshots can be reconsidered as central galaxies again, which is not the case for GP14 and GRP. These latter models assume that once a galaxy becomes a satellite it will remain as such until it merges with a central galaxy \citep{Guo+2016}.
In order to assess the role of satellites which are reconsidered as central galaxies, we remove from the primary sample the central galaxies that were ejected from their host haloes in the past in G13. The results are shown in Table \ref{table_isocentralnoejected}.
Compared to the results shown in Table \ref{table_isocen}, we find that the cumulative S/N of two-halo conformity decreases, down to a 60 percent for the low-mass primaries and 40 percent for the primaries of intermediate masses. The cumulative S/N barely decreases (3 percent) for the massive primaries. Furthermore, the ratio $f_{Q,Q prim}/f_{Q,SF prim}$ at 5 \mpc{} decreases from 6\% to $\leq$4\% for low-mass primaries. 
Therefore, central galaxies ejected from their host halo are partially responsible for the high signal of two-halo conformity found in G13, especially for low-mass primaries.

\begin{table}
\caption{
Same as Table \ref{table_iso}, but for isolated, central galaxies. We do not include central galaxies of G13 that were ejected from their host halo in the past in the sample of primary galaxies.
}
\begin{tabular}{c c c c}
\hline
\hline
log(\mstell) & $\rm S/N$($r<0.5$) & $\rm S/N$($0.5<r<20$) & ratio $f_Q(5)$\\
\hline
G13 & & & \\
\hline
9.5-10  &  19.7  &  50.1  &  1.037 \\
10-10.5  &  66.6  &  42.4  &  1.031 \\
10.5-11   &  74.1  &  49.7  &  1.042 \\
\hline
\end{tabular}
\label{table_isocentralnoejected}
\end{table}

\begin{figure*}
\epsfysize=12cm 
\epsfbox{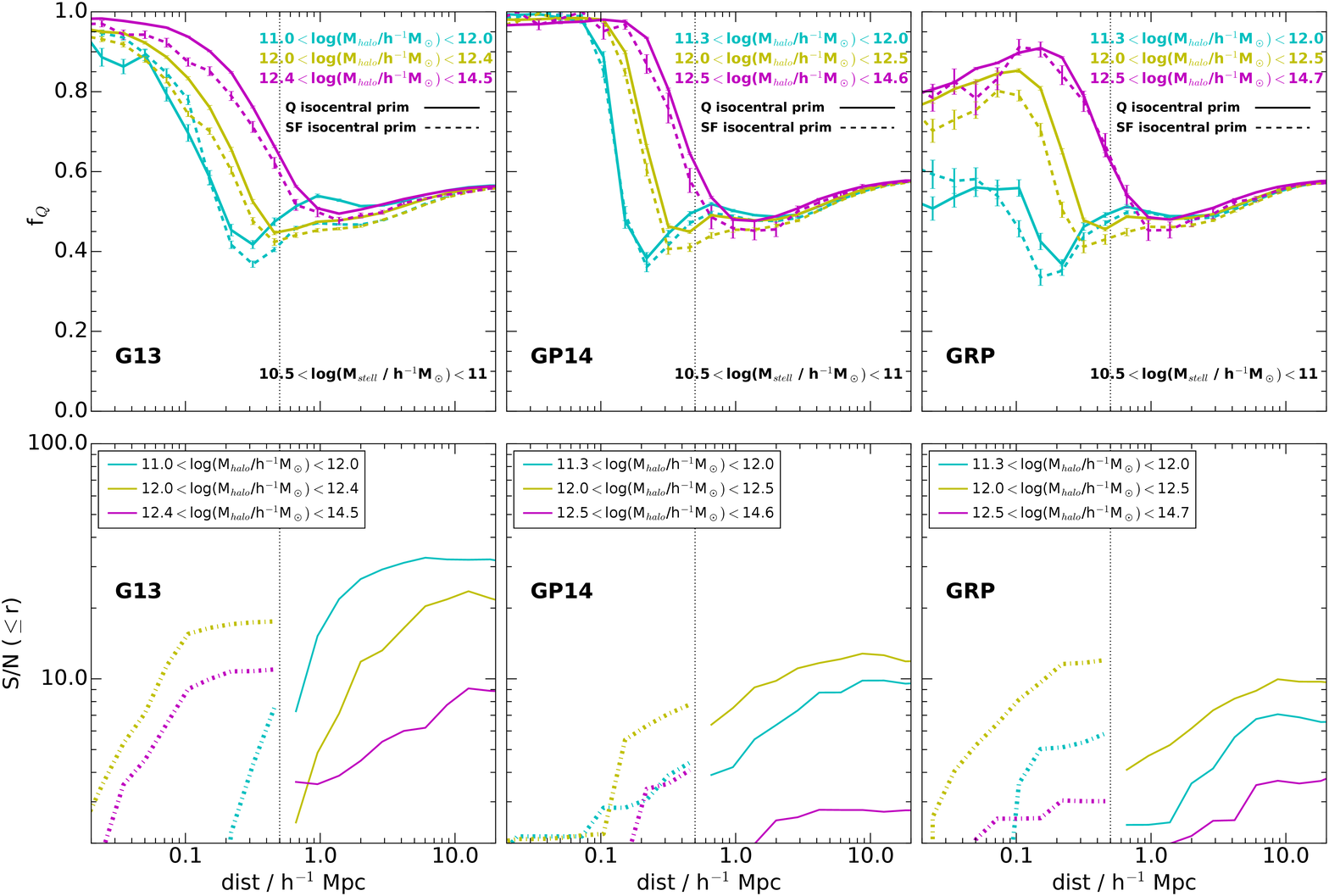}
\caption{
Top panels: mean quenched fraction of secondary galaxies as a function of the distance around Q (solid lines) and SF (dashed lines) primary galaxies using the SAMs of G13 (left-hand panels), GP14 (middle panels) and GRP (right-hand panels). Here the primary galaxies correspond to massive isolated, central galaxies ($10^{10.5} < \mstell/\msun < 10^{11}$). 
The colours correspond to different halo mass bins of primary galaxies indicated in the legend for each SAM. These bins were selected in order to have the same percentage of primary galaxies in each halo mass range.
Bottom panels: cumulative signal-to-noise ratio of galactic conformity
for the cases shown in the top panels.
This is measured separately for the one-halo conformity 
out to $\sim 0.5$ \mpc\ (dash-dot lines) and the two-halo conformity in the range $0.5 < \textrm{dist} < 20$ \mpc\ (solid lines).
The isolation radius is shown as a dotted-vertical line in each panel, which also illustrates our scale to separate the one-halo conformity and the two-halo conformity.
The colours correspond to different halo mass bins of primary galaxies indicated in the legend.
}
\label{plot_PrimMhalo_isocentral}
\end{figure*}

On the other hand, the galactic conformity measured at Mpc scales
could be related with the large scatter in the distribution of halo masses at fixed stellar mass  for the primary galaxies \citep{Paranjape+2015}. 
Fig. \ref{plot_MhMs} in Appendix \ref{app_MhMs} shows the median halo mass as a function of stellar mass ranges of Q and SF isolated,
central primary galaxies for the three SAMs. The scatter in halo mass increases with stellar mass. Furthermore, the median halo mass of Q isolated, central primaries (squares) is higher than that of SF isolated, central primaries (triangles) at fixed stellar mass with \mstell{} $>$ 10$^{10}$ $\msun$ in all the models.
For this reason, we measure the conformity between massive primary galaxies and their neighbours in three different halo mass ranges for the primaries. We use the case of massive primaries since they still show a considerable two-halo conformity using central galaxies in the sample of primaries in G13 (e.g. 4\% in the $f_Q$ ratio at 5 \mpc, value that does not change if we do not include central galaxies that were ejected from their host haloes in the past). For completeness, we also estimate the mean quenched fractions for GP14 and GRP simulations. 
Primary galaxies are all the isolated, central galaxies in G13, GP14 and GRP at $10^{10.5} < \mstell/\msun < 10^{11}$, i.e. we include central galaxies that were ejected in the past in G13.
The halo mass ranges are selected in order to have the same percentage of massive primary galaxies in each halo mass range. The results are shown in Fig. \ref{plot_PrimMhalo_isocentral}. The top panels show the mean quenched fractions around massive Q primaries and massive SF primaries of fixed stellar and halo mass (solid and dashed lines, respectively). The bottom panels show the cumulative S/N for each case.
At a given scale, the overall conformity signal decreases notably compared with that in the bottom panels of Fig. \ref{plot_SNR} (green lines) for each SAM.

The two-halo conformity for the case of massive primaries 
in massive host haloes (\mh $> 10^{12.4}$ $\msun$)
decreases significantly for all the SAMs. In Table \ref{table_Mhalo} we report that the cumulative S/N is $<10$ at 20 \mpc{} and, also, the $f_Q$ ratio is $<$ 2\% at 5 \mpc.
For massive primary galaxies in haloes of lower mass,
S/N($0.5<r<20$) $\lesssim$ 30 in G13, which is smaller compared to that without dividing in halo mass ($\sim$50). 
These results support the claim that the two-halo conformity measured for primary galaxies of the same stellar mass is influenced by primaries residing in haloes of different masses. However, we still detect a significant cumulative signal of two-halo conformity of $\sim 20-30$ between massive isolated, central galaxies in relatively low-mass haloes and their neighbour galaxies in G13.

\begin{table}
\caption{
Conformity signal between massive isolated, central primary galaxies ($10^{10.5} < \mstell/\msun < 10^{11}$) and secondaries (Fig. \ref{plot_PrimMhalo_isocentral}). First column is the halo mass range for the primary galaxies, the second column is the cumulative S/N
out to 0.5 \mpc, and the third column is
the cumulative S/N measured from 0.5 \mpc\ out to 20 \mpc, S/N($0.5<r<20$). The last column
is the ratio between the mean quenched fraction of secondaries around Q and SF massive isolated, central primaries ($f_{Q,Q prim}/f_{Q,SF prim}$) at 5 \mpc.
The halo mass is log$_{10}$ in units of $\msun$.
}
\begin{tabular}{c c c c}
\hline
\hline
log(\mh) & $\rm S/N$($r<0.5$) & $\rm S/N$($0.5<r<20$) & ratio $f_Q(5)$\\
\hline
G13 & & & \\
\hline
11-12  &  7.5  &  32.2  &  1.052 \\
12-12.4  &  17.5  &  22.0  &  1.031 \\
12.4-14.5  &  11.0  &  8.9  &  1.019 \\

\hline
GP14 & & & \\
\hline
11.3-12  &  4.4  &  9.5  &  1.014 \\
12-12.5  &  7.8  &  11.9  &  1.019 \\
12.5-14.6  &  4.1  &  2.8  &  1.014 \\
\hline
GRP & & & \\
\hline
11.3-12  &  5.9  &  6.6  &  1.011 \\
12-12.5  &  12.0  &  9.7  &  1.021 \\
12.5-14.7  &  3.0  &  3.7  &  1.019 \\
\hline
\end{tabular}
\label{table_Mhalo}
\end{table}

In the case of GP14 and GRP, for a given halo mass, the cumulative S/N in both the one-halo and the two-halo conformity regimes is smaller than 12 with $f_Q$ ratios $\lesssim$ 2\% at 5 \mpc{} (see Table \ref{table_Mhalo}).
GP14 and GRP show similar mean quenched fractions around massive primaries at scales beyond the virial radius of the primaries
(middle top and right top panels of Fig. \ref{plot_PrimMhalo_isocentral}, respectively). Therefore, the two-halo conformity is in general independent of the 
prescriptions for satellites used in the SAMs.

\subsection{Conformity between primaries and central galaxies}
\label{sec_censec}

\begin{figure*}
\epsfysize=12cm 
\epsfbox{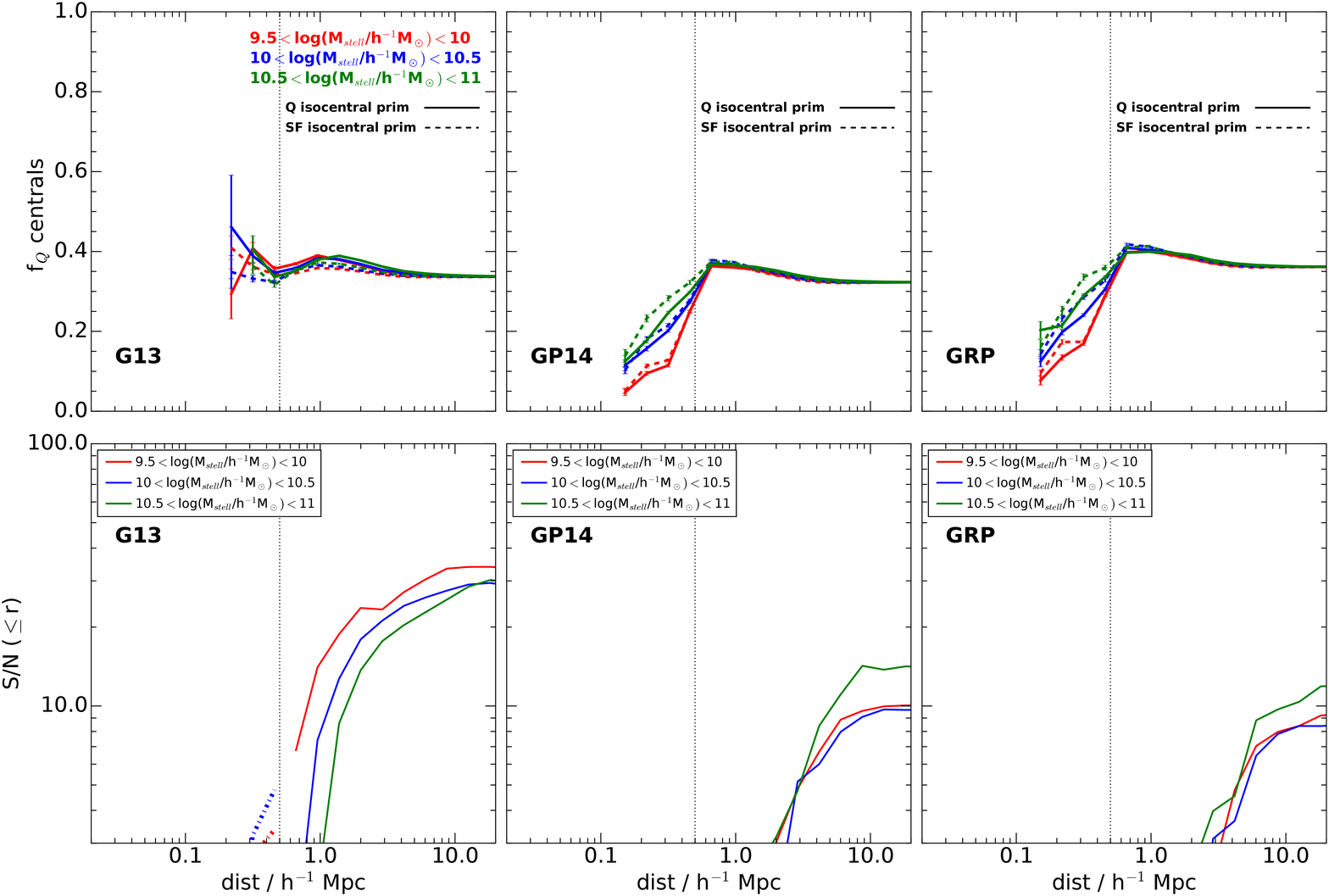}
\caption{
Top panels: mean quenched fraction of central galaxies as a function of the distance around Q (solid lines) and SF (dashed lines) primary galaxies using the SAMs of G13 (left-hand panels), GP14 (middle panels) and GRP (right-hand panels). Here the primary galaxies are selected according to an isolation criterion in real space with a radius of 
0.5 \mpc\ and, in addition, they are 
central galaxies in the SAMs.
We do not include central galaxies that were ejected in the past in G13.
The colours correspond to different stellar mass bins of primary galaxies indicated in the legend.
Bottom panels: cumulative signal-to-noise ratio of galactic conformity
for the cases shown in the top panels.
This is measured separately for the one-halo conformity 
out to $\sim 0.5$ \mpc\ (dash-dot lines) and the two-halo conformity in the range $0.5 < \textrm{dist} < 20$ \mpc\ (solid lines).
The isolation radius is shown as a dotted-vertical line in each panel, which also illustrates our scale to separate the one-halo conformity and the two-halo conformity.
}
\label{plot_Prim_isocentral_Sec_central}
\end{figure*}

In this section we explore the case of galactic conformity between isolated, central galaxies and central galaxies, i.e. we do not include satellite galaxies in the sample of secondary galaxies.
In addition, we do not include central galaxies that were ejected from their host halo in the past in G13 since these galaxies increase the signal of two-halo conformity compared with GRP and GP14 models as shown in the previous section.
Fig. \ref{plot_Prim_isocentral_Sec_central} shows the results. The top panels correspond to the mean quenched fraction of central galaxies around isolated, central galaxies for the three SAMs. Of course, this fraction is zero at small scales because we do not include satellite galaxies. Other central galaxies start to appear in the infall region of the primary galaxies (typically between 0.1 and 0.3 \mpc). At larger scales, the conformity is rather weak for G13, GP14 and GRP. 
For all the SAMs, the quenched fractions decrease monotonically at Mpc scales since we do not consider satellite galaxies in other haloes (c.f. Fig.\ref{plot_Prim_IsoCentral}).

As can be seen in the bottom panels of Fig. \ref{plot_Prim_isocentral_Sec_central}, the cumulative S/N 
of two-halo conformity is higher
for G13. This value gets up to $\sim 30$--35 around primaries at 20 \mpc, whereas the cumulative S/N is always lower than 15 around primaries of all the masses for GP14 and GRP. 
The overall cumulative signal of central galaxies as secondaries is lower than that using all the neighbour galaxies (centrals and satellites) as secondaries, especially for G13.

\section{Discussion}
\label{sec_discussion}

Our finding of one-halo galactic conformity in the semi-analytic models of G13 and GRP between isolated primary galaxies and their neighbour galaxies (Section \ref{sec_results_isoprim}) are consistent with the galactic conformity observed in galaxy groups by other authors \citep[e.g.][]{WangWhite2012,Phillips+2014}.
That is, the fraction of quenched neighbour galaxies at scales smaller than $\sim$0.3 \mpc\ is higher for quenched primary galaxies than for star-forming primaries at fixed stellar mass. This shows that SAMs  
are able to reproduce the observed correlations between central and satellite galaxies. The signal of this correlation depends on the specific model. 
In the case of GP14, the 
environmental process of removing, instantaneously, all hot gas in satellites
is so strong that nearly all the neighbour galaxies are quenched inside the virial radius of the central galaxy. 
Interestingly, a galactic conformity appears for scales larger than 0.1 \mpc\ in this model. 

As a matter of fact, the three models have high cumulative S/N of two-halo conformity at scales $> 0.5$ \mpc. Therefore, regardless the feedback acting inside the virial radius of central galaxies, a correlation between quenched (star-forming) central galaxies and quenched (star-forming) neighbour galaxies is detected within scales that correspond to the infall region of haloes and beyond. As well as the case of one-halo conformity, the amplitude of this signal is strongly dependent on each SAM. 
G13 shows the highest two-halo conformity.
The three SAMs consistently show  low-mass isolated primary galaxies and their neighbours  with  higher cumulative S/N of conformity at Mpc scales.

Some authors have reported a two-halo conformity in observations \citep[e.g.][]{Kauffmann+2013,Kauffmann2015,Berti+2017}.
The chosen criterion of isolated primaries in this work is similar to that used in observations that look for the conformity signal, but we use the 3D information provided by the models to estimate the distance. 
We expect the signal of conformity to be diluted if the isolation criterion is applied in redshift space.
We confirm that the isolation criterion allows to add some satellite galaxies in the sample of primary galaxies \citep[see][]{Kauffmann+2013,Bray+2016,Sin+2017,Tinker+2017}. 
Although the fraction of satellites in the primary sample is relatively small ($\leq$ 6\% for the low-mass primary galaxies), they contribute to increase the cumulative S/N of conformity compared to the case when we use isolated galaxies that are true central galaxies in the SAMs, especially for the low-mass primary galaxies (see Fig. \ref{plot_SNR}). 
This is consistent with \citet{Tinker+2017} and \citet{Sin+2017} where the large-scale galactic conformity notoriously decreases after removing a small fraction of satellite galaxies in the primary sample (6.5\% and 7\%, respectively, for masses between $10^{10}$ and $10^{10.5}$ M$_{\bigodot}$) using their group catalogs. Around two-thirds of these satellites are located near large galaxy clusters \citep{Sin+2017}.

Two-halo conformity at scales larger than 1 \mpc\ is 
low
regardless of the mass of the isolated, central primary galaxies for all the SAMs studied here, except G13. The fact that GP14 and GRP show similar results at large scales although they have different 
treatment of the hot gas in satellites
suggests that other  
effects are playing a role in the two-halo conformity. 
We removed from the primary sample the central galaxies that were ejected in the past in G13 and found that in this case the overall cumulative S/N decreases, down to a 60 percent for the low-mass primaries. 
This result shows that central galaxies ejected from their host halo are partially responsible of the higher signal of two-halo conformity measured in G13 compared with GP14 and GRP, especially for low-mass primary galaxies.
 
We also explored the case of conformity between massive isolated, central primary galaxies and their neighbour galaxies at both fixed stellar mass and fixed halo mass for the primaries (Fig. \ref{plot_PrimMhalo_isocentral}). The two-halo conformity is detected 
for massive primaries in relatively low-mass haloes ($\mh$ $< 10^{12.4}$ $\msun$) in G13. In this case, the cumulative S/N is weaker (a factor of two) compared to the case of fixing only the stellar mass for the primaries. 
Our result is qualitatively similar to that of \citet{Bray+2016}. They find that the mean red fraction of secondaries around redder isolated galaxies is higher than around blue isolated galaxies at fixed halo mass ($\mh$ $< 10^{12}$ $\msun$) out to Mpc scales in the Illustris simulation. 
They use an isolation radius of 0.35 \mpc\ (0.5 Mpc) in real space.
The difference in the mean red fractions is similar to that at fixed stellar mass for the primaries which leads them to conclude that using stellar mass as a proxy for halo mass is unlikely to be biased by this selection technique. However, we note that this would not be valid for central galaxies in massive haloes since our results show that the cumulative S/N is $< 10$ for massive primaries in haloes with \mh $>$ 10$^{12.4}$ $\msun$. 
The signal of two-halo conformity is influenced 
by primary galaxies at fixed stellar mass that reside in dark matter haloes of different masses.

We also estimated the mean quenched fraction of central galaxies around isolated, central primaries (Fig. \ref{plot_Prim_isocentral_Sec_central}). In G13 we do not include central galaxies that were ejected from their host halo in the past. The cumulative S/N reaches $\sim$35 for the low-mass primaries, whereas this value is lower than 15 for both GP14 and GRP. 
\citet{Bray+2016} studied a similar case in Illustris using colour instead of sSFR. They found that the signal of conformity out to $\sim$ 3 Mpc 
is still present for low-mass primaries, and nearly identical to the signal in the case with all the neighbour galaxies (i.e. satellites and central) in the secondary sample, but the signal between 3 and 10 Mpc is completely suppressed. 
This is similar to our result using G13, although we find that the cumulative signal of conformity at 20 \mpc\ is reduced by a 30\% when we do not consider satellite galaxies in the sample of secondaries.
In the case of GP14 and GRP, the cumulative S/N is lower at all the scales when the satellites are not included as secondary galaxies. 

Since in the SAMs the central galaxies are tracing the position of their host haloes, the results shown in Fig. \ref{plot_Prim_isocentral_Sec_central} can be thought as an attempt to measure the conformity between haloes. 
If we assume that stellar mass is a proxy of halo mass and sSFR is a proxy of some physical property of the haloes related to their formation time, this measurement can serve as a link between conformity and the assembly bias. The results using G13  show that conformity is smaller than 3\% 
beyond $\sim 3$ \mpc\ for all the masses in this model. 
Furthermore, in the case of GP14 and GRP, the signal of the proxy of conformity between haloes is very weak ($<$1--2\%) at all the scales and masses. 
This is consistent with \citealt{Bray+2016} and \citealt{Tinker+2017} where conformity is not present at scales larger than 3 \mpc.
It seems difficult that galactic conformity is directly related with a long-range effect, otherwise we should measure a strong growth in the cumulative S/N of conformity at least between 5 and 10 \mpc, which is not the case.
On the other hand, \citet{Sin+2017} have found that the conformity out to projected distances of 4 Mpc 
is primarily related with the environmental influence of large galaxy clusters on central galaxies in high-density environments. \citet{Lacerna2011} found that the assembly bias effect can be explained by
old, small structures located near massive haloes that are typically at distances out to 4 virial radii ($\lesssim 1.5$ \mpc). 
These massive haloes could disrupt the normal growth of near small objects and, therefore, affect their virial masses and ages.
A more in depth analysis of whether conformity measured in models out to $\sim$ 3 \mpc\ is reflecting assembly bias will be present elsewhere.

\section{Conclusions}
\label{conclusions}

We have studied the correlation between the sSFR of central galaxies and neighbour galaxies out to several Mpc scales using three SAMs (G13, GP14, and GRP). For that, we measure the mean quenched fraction of 
secondary galaxies around 
primary galaxies, 
where the latter are intended to be central galaxies.
The secondary galaxies correspond to all the galaxies in the vicinity of primary galaxies.
In all the models we find that the mean quenched fraction at distances larger than 0.5 \mpc\ around quenched primary galaxies is higher than that around star-forming primary galaxies of the same stellar mass, i.e. we detect two-halo galactic conformity, when the selection of primary galaxies is based on an isolation criterion as in observations so as to select central galaxies, but in real space. The signal of two-halo conformity, measured as the cumulative S/N, decreases with the stellar mass of the primary galaxies in this case.
For a given stellar mass of the primaries, the cumulative signal 
of conformity depends strongly on each SAM. The G13 model shows the higher cumulative S/N 
of two-halo conformity. 
For the low-mass primaries ($10^{9.5} < \mstell/\msun < 10^{10}$), the cumulative S/N of two-halo conformity is $\sim 145$ at 20 \mpc\ and  the  
the mean quenched fraction of secondaries around Q isolated primaries is 7\% higher than that around SF isolated primaries
at 5 \mpc\ in G13.
In contrast, at the same stellar mass, the cumulative S/N of two-halo conformity is $<$ 100
at 20 \mpc\ in the GP14 and GRP models, with differences in the mean quenched fraction of secondaries around Q and SF primaries at 5 \mpc\ smaller than 3\%. 

The isolation criterion includes a small fraction of satellites galaxies in the sample of primary galaxies ($\lesssim 6\%$ for low-mass primaries). The overall galactic conformity decreases when we add the condition of central galaxies, according to the information given by the SAMs, in the selection of the primaries. The cumulative S/N of two-halo conformity decreases dramatically down to $\lesssim$ 20 in the GP14 and GRP models for all the masses of the primary galaxies in this case. The differences in the mean quenched fraction of secondaries around Q and SF primaries at 5 \mpc\ are smaller than 1\% for low-mass primaries and primaries of intermediate masses, whereas are smaller than 3\% for massive primaries.
For the G13 model, the cumulative S/N is still $>$ 100 at 20 \mpc\
for low-mass primaries and with differences between 4\% and 6\% in the mean quenched fraction of secondaries around Q and SF primaries at 5 \mpc\ at all the masses studied.

Since 
in GP14 and GRP once a galaxy becomes a satellite remains as such, whereas in G13 galaxies that are satellites can become centrals at a later time,
we assess the role of satellites which are reconsidered later as central galaxies by removing from the primary sample the central galaxies that were ejected from their host halo in the past in G13. 
We find that the cumulative S/N of two-halo conformity is $\lesssim$ 50. It decreases down to a 60\% for the low-mass primaries and 40\% percent for the primaries of intermediate masses compared with the case of  isolated, central galaxies described above. Therefore, central galaxies that were ejected from their host halo in the past are partially responsible of the high signal of two-halo conformity measured in G13, especially for low-mass primaries.

We then explore if the two-halo conformity in G13 is also related with the scatter in the distribution of halo masses at fixed stellar mass. We use massive primaries for which the cumulative S/N of two-halo conformity barely decreased (3\%)
after removing the ejected (massive) centrals. We find that the two-halo conformity is only detected 
for isolated, central galaxies in relatively low-mass haloes (\mh $< 10^{12.4}$ $\msun$) with a cumulative S/N of $\sim$20 -- 30.
Our results support that the two-halo conformity is also influenced by primary galaxies at fixed stellar mass that reside in dark matter haloes of different masses.

Finally, we explore the case of conformity when the secondary galaxies correspond to central galaxies in the SAMs as an attempt to measure the correlation of conformity between distinct haloes. Our results show that the two-halo conformity is weak beyond $\sim 3$ \mpc\ for all the mass ranges in the SAMs ($<$3\% in G13, $<$1--2\% in both GP14 and GRP models).
Therefore, it seems difficult that conformity is directly related with a long-range effect.
It is likely that the conformity measured at scales smaller than 3--4 \mpc\ is related with the environmental influence of large galaxy clusters on central galaxies hosted by lower mass haloes.

\section*{Acknowledgments}
We thank the anonymous referee for his/her constructive revision of our paper.
NP \& SC acknowledge support from a STFC/Newton Fund award (ST/M007995/1 - DPI20140114). 
SC further acknowledges support from CONICYT Doctoral Fellowship Programme. 
NP was supported by Fondecyt Regular 1150300 and Anillo ACT-1417.
VGP acknowledges support from the University of Portsmouth through the Dennis Sciama Fellowship award. 
Part of these calculations were performed using the Geryon cluster at the Center
for Astro-Engineering UC, part of the BASAL CATA PFB-06, which received joint funding from Anillo ACT-86, FONDEQUIP AIC-57 and QUIMAL 130008.
This work also used the DiRAC Data Centric system at Durham University, operated by the Institute for Computational Cosmology on behalf of the STFC DiRAC HPC Facility (www.dirac.ac.uk). This equipment was funded by BIS National E-infrastructure capital grant ST/K00042X/1, STFC capital grants ST/H008519/1 and ST/K00087X/1, STFC DiRAC Operations grant ST/K003267/1 and Durham University. DiRAC is part of the National E-Infrastructure. This work has benefited from the publicly available programming language {\sc python}.
\bibliography{references}

\begin{thebibliography}{75}
\providecommand{\natexlab}[1]{#1}

\bibitem[{{Abazajian} et~al.(2004)}]{Abazajian:2004}
{Abazajian} K. et~al., 2004, \aj, 128, 502

\bibitem[{{Abazajian} et~al.(2009)}]{Abazajian:2009}
{Abazajian} K.~N. et~al., 2009, \apjs, 182, 543

\bibitem[{{Aihara} et~al.(2011)}]{Aihara+2011}
{Aihara} H. et~al., 2011, \apjs, 193, 29

\bibitem[{{Alam} et~al.(2015)}]{Alam+2015}
{Alam} S. et~al., 2015, \apjs, 219, 12

\bibitem[{{Angulo} et~al.(2008){Angulo}, {Baugh} \& {Lacey}}]{Angulo+2008b}
{Angulo} R.~E., {Baugh} C.~M., {Lacey} C.~G., 2008, \mnras, 387, 921

\bibitem[{{Berti} et~al.(2017){Berti}, {Coil}, {Behroozi}, {Eisenstein},
  {Bray}, {Cool} \& {Moustakas}}]{Berti+2017}
{Berti} A.~M., {Coil} A.~L., {Behroozi} P.~S., {Eisenstein} D.~J., {Bray}
  A.~D., {Cool} R.~J., {Moustakas} J., 2017, \apj, 834, 87

\bibitem[{{Bett} et~al.(2007){Bett}, {Eke}, {Frenk}, {Jenkins}, {Helly} \&
  {Navarro}}]{Bett+2007}
{Bett} P., {Eke} V., {Frenk} C.~S., {Jenkins} A., {Helly} J., {Navarro} J.,
  2007, \mnras, 376, 215

\bibitem[{{Blanton} et~al.(2005)}]{Blanton+2005}
{Blanton} M.~R. et~al., 2005, \aj, 129, 2562

\bibitem[{{Bower} et~al.(2006)}]{Bower:2006}
{Bower} R.~G., {Benson} A.~J., {Malbon} R., {Helly} J.~C., {Frenk} C.~S.,
  {Baugh} C.~M., {Cole} S., {Lacey} C.~G., 2006, \mnras, 370, 645

\bibitem[{{Boylan-Kolchin} et~al.(2009){Boylan-Kolchin}, {Springel}, {White},
  {Jenkins} \& {Lemson}}]{Boylan-Kolchin+2009}
{Boylan-Kolchin} M., {Springel} V., {White} S.~D.~M., {Jenkins} A., {Lemson}
  G., 2009, \mnras, 398, 1150

\bibitem[{{Bray} et~al.(2016)}]{Bray+2016}
{Bray} A.~D. et~al., 2016, \mnras, 455, 185

\bibitem[{{Campbell} et~al.(2015)}]{Campbell+2015}
{Campbell} D., {van den Bosch} F.~C., {Hearin} A., {Padmanabhan} N., {Berlind}
  A., {Mo} H.~J., {Tinker} J., {Yang} X., 2015, \mnras, 452, 444

\bibitem[{{Coil} et~al.(2011)}]{Coil+2011}
{Coil} A.~L. et~al., 2011, \apj, 741, 8

\bibitem[{{Cole} et~al.(2000){Cole}, {Lacey}, {Baugh} \& {Frenk}}]{Cole:2000}
{Cole} S., {Lacey} C.~G., {Baugh} C.~M., {Frenk} C.~S., 2000, \mnras, 319, 168

\bibitem[{{Contreras} et~al.(2013){Contreras}, {Baugh}, {Norberg} \&
  {Padilla}}]{Contreras+2013}
{Contreras} S., {Baugh} C.~M., {Norberg} P., {Padilla} N., 2013, \mnras, 432,
  2717

\bibitem[{{Cool} et~al.(2013)}]{Cool+2013}
{Cool} R.~J. et~al., 2013, \apj, 767, 118

\bibitem[{{Croton} et~al.(2006)}]{Croton:2006}
{Croton} D.~J. et~al., 2006, \mnras, 365, 11

\bibitem[{{Dav{\'e}} et~al.(2016){Dav{\'e}}, {Thompson} \&
  {Hopkins}}]{Dave+2016}
{Dav{\'e}} R., {Thompson} R., {Hopkins} P.~F., 2016, \mnras, 462, 3265

\bibitem[{{Davis} et~al.(1985){Davis}, {Efstathiou}, {Frenk} \&
  {White}}]{Davis+1985}
{Davis} M., {Efstathiou} G., {Frenk} C.~S., {White} S.~D.~M., 1985, \apj, 292,
  371

\bibitem[{{De Lucia} \& {Blaizot}(2007)}]{DeLucia:2007}
{De Lucia} G., {Blaizot} J., 2007, \mnras, 375, 2

\bibitem[{{De Lucia} et~al.(2004){De Lucia}, {Kauffmann} \&
  {White}}]{DeLucia:2004}
{De Lucia} G., {Kauffmann} G., {White} S.~D.~M., 2004, \mnras, 349, 1101

\bibitem[{{Faltenbacher} \& {White}(2010)}]{FW10}
{Faltenbacher} A., {White} S.~D.~M., 2010, \apj, 708, 469

\bibitem[{{Font} et~al.(2008)}]{Font:2008}
{Font} A.~S. et~al., 2008, \mnras, 289, 1619

\bibitem[{{Gao} \& {White}(2007)}]{Gao-White07}
{Gao} L., {White} S.~D.~M., 2007, \mnras, 377, L5

\bibitem[{{Gao} et~al.(2005){Gao}, {Springel} \& {White}}]{Gao05}
{Gao} L., {Springel} V., {White} S.~D.~M., 2005, \mnras, 363, L66

\bibitem[{{Gonzalez-Perez} et~al.(2014){Gonzalez-Perez}, {Lacey}, {Baugh},
  {Lagos}, {Helly}, {Campbell} \& {Mitchell}}]{Gonzalez-Perez:2014}
{Gonzalez-Perez} V., {Lacey} C.~G., {Baugh} C.~M., {Lagos} C.~D.~P., {Helly}
  J., {Campbell} D.~J.~R., {Mitchell} P.~D., 2014, \mnras, 439, 264

\bibitem[{{Granato} et~al.(2000){Granato}, {Lacey}, {Silva}, {Bressan},
  {Baugh}, {Cole} \& {Frenk}}]{granato00}
{Granato} G.~L., {Lacey} C.~G., {Silva} L., {Bressan} A., {Baugh} C.~M., {Cole}
  S., {Frenk} C.~S., 2000, \apj, 542, 710

\bibitem[{{Guo} et~al.(2011)}]{Guo:2011}
{Guo} Q. et~al., 2011, \mnras, 413, 101

\bibitem[{{Guo} et~al.(2013)}]{Guo:2013}
{Guo} Q., {White} S., {Angulo} R.~E., {Henriques} B., {Lemson} G.,
  {Boylan-Kolchin} M., {Thomas} P., {Short} C., 2013, \mnras, 428, 1351

\bibitem[{{Guo} et~al.(2016)}]{Guo+2016}
{Guo} Q. et~al., 2016, \mnras, 461, 3457

\bibitem[{{Hartley} et~al.(2015){Hartley}, {Conselice}, {Mortlock}, {Foucaud}
  \& {Simpson}}]{Hartley+2015}
{Hartley} W.~G., {Conselice} C.~J., {Mortlock} A., {Foucaud} S., {Simpson} C.,
  2015, \mnras, 451, 1613

\bibitem[{{Hearin}(2015)}]{Hearin2015}
{Hearin} A.~P., 2015, \mnras, 451, L45

\bibitem[{{Hearin} et~al.(2015){Hearin}, {Watson} \& {van den
  Bosch}}]{Hearin+2015}
{Hearin} A.~P., {Watson} D.~F., {van den Bosch} F.~C., 2015, \mnras, 452, 1958

\bibitem[{{Hearin} et~al.(2016){Hearin}, {Behroozi} \& {van den
  Bosch}}]{Hearin+2016}
{Hearin} A.~P., {Behroozi} P.~S., {van den Bosch} F.~C., 2016, \mnras, 461,
  2135

\bibitem[{{Henriques} et~al.(2013){Henriques}, {White}, {Thomas}, {Angulo},
  {Guo}, {Lemson} \& {Springel}}]{Henriques:2013}
{Henriques} B.~M.~B., {White} S.~D.~M., {Thomas} P.~A., {Angulo} R.~E., {Guo}
  Q., {Lemson} G., {Springel} V., 2013, \mnras, 431, 3373

\bibitem[{{Henriques} et~al.(2015)}]{Henriques:2015}
{Henriques} B.~M.~B., {White} S.~D.~M., {Thomas} P.~A., {Angulo} R., {Guo} Q.,
  {Lemson} G., {Springel} V., {Overzier} R., 2015, \mnras, 451, 2663

\bibitem[{{Jiang} et~al.(2014){Jiang}, {Helly}, {Cole} \& {Frenk}}]{Jiang:2014}
{Jiang} L., {Helly} J.~C., {Cole} S., {Frenk} C.~S., 2014, \mnras, 440, 2115

\bibitem[{{Kauffmann}(2015)}]{Kauffmann2015}
{Kauffmann} G., 2015, \mnras, 454, 1840

\bibitem[{{Kauffmann} et~al.(2013){Kauffmann}, {Li}, {Zhang} \&
  {Weinmann}}]{Kauffmann+2013}
{Kauffmann} G., {Li} C., {Zhang} W., {Weinmann} S., 2013, \mnras, 430, 1447

\bibitem[{{Kawinwanichakij} et~al.(2016)}]{Kawinwanichakij+2016}
{Kawinwanichakij} L. et~al., 2016, \apj, 817, 9

\bibitem[{{Kennicutt}(1998)}]{ken98}
{Kennicutt} Jr. R.~C., 1998, \apj, 498, 541

\bibitem[{{Kerscher}(2017)}]{Kerscher2017}
{Kerscher} M., 2017, preprint (arXiv:1705.07582)

\bibitem[{{Knobel} et~al.(2015){Knobel}, {Lilly}, {Woo} \& {Kova{\v
  c}}}]{Knobel+2015}
{Knobel} C., {Lilly} S.~J., {Woo} J., {Kova{\v c}} K., 2015, \apj, 800, 24

\bibitem[{{Komatsu} et~al.(2011)}]{Komatsu+2011}
{Komatsu} E. et~al., 2011, \apjs, 192, 18

\bibitem[{{Lacerna} \& {Padilla}(2011)}]{Lacerna2011}
{Lacerna} I., {Padilla} N., 2011, \mnras, 412, 1283

\bibitem[{{Lacerna} \& {Padilla}(2012)}]{Lacerna2012}
{Lacerna} I., {Padilla} N., 2012, \mnras, 426, L26

\bibitem[{{Lacey} et~al.(2016)}]{Lacey:2016}
{Lacey} C.~G. et~al., 2016, \mnras, 462, 3854

\bibitem[{{Lagos} et~al.(2011){Lagos}, {Lacey}, {Baugh}, {Bower} \&
  {Benson}}]{Lagos:2011a}
{Lagos} C.~D.~P., {Lacey} C.~G., {Baugh} C.~M., {Bower} R.~G., {Benson} A.~J.,
  2011, \mnras, 416, 1566

\bibitem[{{Lagos} et~al.(2012){Lagos}, {Bayet}, {Baugh}, {Lacey}, {Bell},
  {Fanidakis} \& {Geach}}]{Lagos:2012}
{Lagos} C.~d.~P., {Bayet} E., {Baugh} C.~M., {Lacey} C.~G., {Bell} T.~A.,
  {Fanidakis} N., {Geach} J.~E., 2012, \mnras, 426, 2142

\bibitem[{{Lagos} et~al.(2014){Lagos}, {Davis}, {Lacey}, {Zwaan}, {Baugh},
  {Gonzalez-Perez} \& {Padilla}}]{Lagos:2014b}
{Lagos} C.~d.~P., {Davis} T.~A., {Lacey} C.~G., {Zwaan} M.~A., {Baugh} C.~M.,
  {Gonzalez-Perez} V., {Padilla} N.~D., 2014, \mnras, 443, 1002

\bibitem[{{Lawrence} et~al.(2007)}]{Lawrence+2007}
{Lawrence} A. et~al., 2007, \mnras, 379, 1599

\bibitem[{{Li} et~al.(2008){Li}, {Mo} \& {Gao}}]{Li+2008}
{Li} Y., {Mo} H.~J., {Gao} L., 2008, \mnras, 389, 1419

\bibitem[{{Lin} et~al.(2016)}]{Lin+2016}
{Lin} Y.~T., {Mandelbaum} R., {Huang} Y.~H., {Huang} H.~J., {Dalal} N.,
  {Diemer} B., {Jian} H.~Y., {Kravtsov} A., 2016, \apj, 819, 119

\bibitem[{{McCracken} et~al.(2012)}]{McCracken+2012}
{McCracken} H.~J. et~al., 2012, \aap, 544, A156

\bibitem[{{More} et~al.(2016)}]{More+2016}
{More} S. et~al., 2016, \apj, 825, 39

\bibitem[{{Norberg} et~al.(2009){Norberg}, {Baugh}, {Gazta{\~n}aga} \&
  {Croton}}]{Norberg09}
{Norberg} P., {Baugh} C.~M., {Gazta{\~n}aga} E., {Croton} D.~J., 2009, \mnras,
  396, 19

\bibitem[{{Paranjape} et~al.(2015){Paranjape}, {Kova{\v c}}, {Hartley} \&
  {Pahwa}}]{Paranjape+2015}
{Paranjape} A., {Kova{\v c}} K., {Hartley} W.~G., {Pahwa} I., 2015, \mnras,
  454, 3030

\bibitem[{{Phillips} et~al.(2014){Phillips}, {Wheeler}, {Boylan-Kolchin},
  {Bullock}, {Cooper} \& {Tollerud}}]{Phillips+2014}
{Phillips} J.~I., {Wheeler} C., {Boylan-Kolchin} M., {Bullock} J.~S., {Cooper}
  M.~C., {Tollerud} E.~J., 2014, \mnras, 437, 1930

\bibitem[{{Rafieferantsoa} \& {Dav{\'e}}(2017)}]{RD+2017}
{Rafieferantsoa} M., {Dav{\'e}} R., 2017, preprint (arXiv:1707.01950)

\bibitem[{{Sin} et~al.(2017){Sin}, {Lilly} \& {Henriques}}]{Sin+2017}
{Sin} L.~P.~T., {Lilly} S.~J., {Henriques} B.~M.~B., 2017, \mnras, 471, 1192

\bibitem[{{Springel}(2005)}]{Springel2005}
{Springel} V., 2005, \mnras, 364, 1105

\bibitem[{{Springel} et~al.(2001){Springel}, {White}, {Tormen} \&
  {Kauffmann}}]{Springel+2001}
{Springel} V., {White} S.~D.~M., {Tormen} G., {Kauffmann} G., 2001, \mnras,
  328, 726

\bibitem[{{Springel} et~al.(2005)}]{Springel+2005}
{Springel} V. et~al., 2005, \nat, 435, 629

\bibitem[{{Tinker} et~al.(2017){Tinker}, {Hahn}, {Mao}, {Wetzel} \&
  {Conroy}}]{Tinker+2017}
{Tinker} J.~L., {Hahn} C., {Mao} Y.~Y., {Wetzel} A.~R., {Conroy} C., 2017,
  preprint (arXiv:1702.01121)

\bibitem[{{Tojeiro} et~al.(2017)}]{Tojeiro+2017}
{Tojeiro} R. et~al., 2017, \mnras, 470, 3720

\bibitem[{{van Daalen} et~al.(2012){van Daalen}, {Angulo} \&
  {White}}]{vanDaalen+2012}
{van Daalen} M.~P., {Angulo} R.~E., {White} S.~D.~M., 2012, \mnras, 424, 2954

\bibitem[{{Vogelsberger} et~al.(2014)}]{Vogelsberger+2014}
{Vogelsberger} M. et~al., 2014, \mnras, 444, 1518

\bibitem[{{Wang} \& {White}(2012)}]{WangWhite2012}
{Wang} W., {White} S.~D.~M., 2012, \mnras, 424, 2574

\bibitem[{{Wechsler} et~al.(2006){Wechsler}, {Zentner}, {Bullock}, {Kravtsov}
  \& {Allgood}}]{Wechsler06}
{Wechsler} R.~H., {Zentner} A.~R., {Bullock} J.~S., {Kravtsov} A.~V., {Allgood}
  B., 2006, \apj, 652, 71

\bibitem[{{Weinmann} et~al.(2006){Weinmann}, {van den Bosch}, {Yang} \&
  {Mo}}]{Weinmann+2006}
{Weinmann} S.~M., {van den Bosch} F.~C., {Yang} X., {Mo} H.~J., 2006, \mnras,
  366, 2

\bibitem[{{Wetzel} et~al.(2007){Wetzel}, {Cohn}, {White}, {Holz} \&
  {Warren}}]{Wetzel+2007}
{Wetzel} A.~R., {Cohn} J.~D., {White} M., {Holz} D.~E., {Warren} M.~S., 2007,
  \apj, 656, 139

\bibitem[{{Zehavi} et~al.(2002)}]{Zehavi02}
{Zehavi} I. et~al., 2002, \apj, 571, 172

\bibitem[{{Zhu} et~al.(2006){Zhu}, {Zheng}, {Lin}, {Jing}, {Kang} \&
  {Gao}}]{Zhu06}
{Zhu} G., {Zheng} Z., {Lin} W.~P., {Jing} Y.~P., {Kang} X., {Gao} L., 2006,
  \apjl, 639, L5

\bibitem[{{Zu} \& {Mandelbaum}(2017)}]{ZM+2017}
{Zu} Y., {Mandelbaum} R., 2017, preprint (arXiv:1703.09219)

\bibitem[{{Zu} et~al.(2017){Zu}, {Mandelbaum}, {Simet}, {Rozo} \&
  {Rykoff}}]{Zu+2017}
{Zu} Y., {Mandelbaum} R., {Simet} M., {Rozo} E., {Rykoff} E.~S., 2017, \mnras,
  470, 551

\end{thebibliography}
\label{lastpage}

\appendix

\section{Halo mass distribution at fixed stellar mass}
\label{app_MhMs}

\begin{figure*}
\epsfysize=6.3cm 
\epsfbox{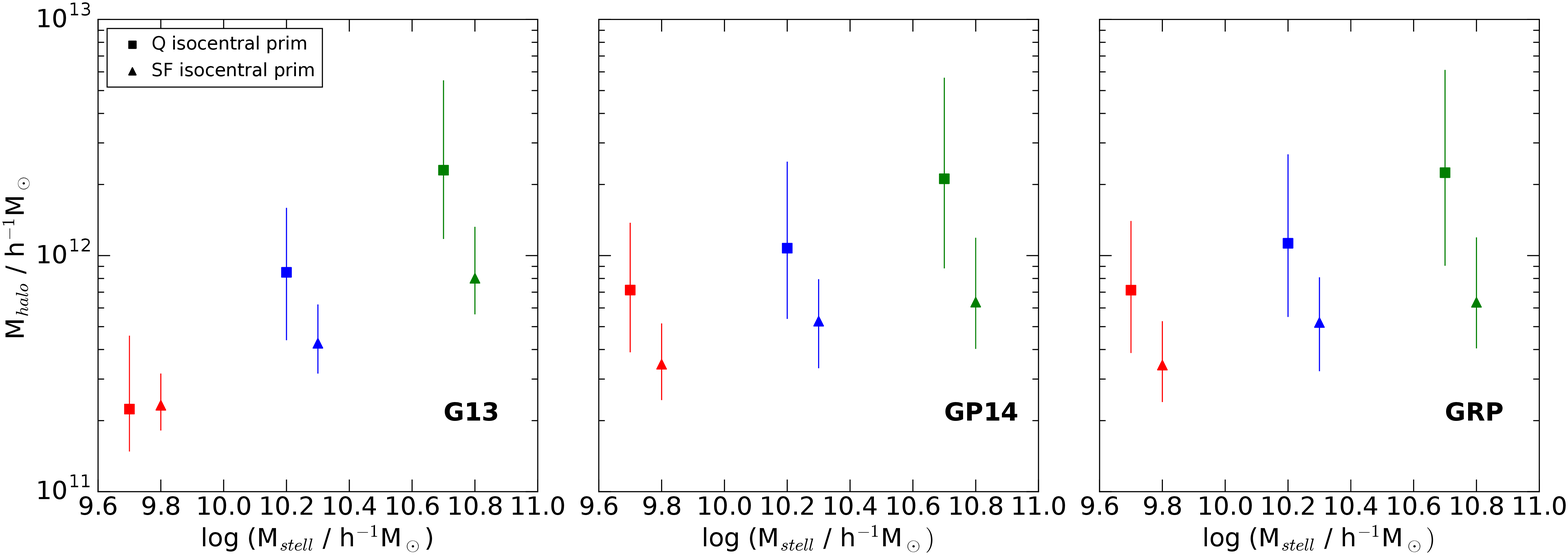}
\caption{
Median halo mass as a function of stellar mass for Q (squares) and SF (triangles) isolated, central primary galaxies using the SAMs of G13 (left-hand panel), GP14 (middle panel) and GRP (right-hand panel). The error bars correspond to the 16th and 84th percentiles of the halo mass distribution. The red, blue, and green symbols correspond to the median halo masses of primary galaxies at 9.5 $<$ log \mstell/$\msun$ $<$ 10, 10 $<$ log \mstell/$\msun$ $<$ 10.5, and 10.5 $<$ log \mstell/$\msun$ $<$ 11, respectively. We apply an offset in stellar mass for each range of \mstell{} between Q and SF primaries to improve clarity.
}
\label{plot_MhMs}
\end{figure*}

Figure \ref{plot_MhMs} shows the median halo mass as a function of stellar mass  for Q and SF isolated, central primary galaxies for the three SAMs. The error bars correspond to the 16th and 84th percentiles of the halo mass distribution. At fixed stellar mass, the scatter in halo mass is comparable among the models. However, the median halo mass of quenched isolated, central primaries (squares) is higher than that of star-forming isolated, central primaries (triangles) at fixed stellar mass with \mstell{} $>$ 10$^{10}$ $\msun$ in all the models.

\end{document}